\documentclass[twocolumn]{article} 
\usepackage{graphicx} 
\usepackage{amssymb}

\long\def\comment#1{}

\makeatletter
\def\@normalsize{\@setsize\normalsize{10pt}\xpt\@xpt
\abovedisplayskip 10pt plus2pt minus5pt\belowdisplayskip
\abovedisplayskip \abovedisplayshortskip \z@
plus3pt\belowdisplayshortskip 6pt plus3pt
minus3pt\let\@listi\@listI}

\def\subsize{\@setsize\subsize{12pt}\xipt\@xipt}
\def\section{\@startsection {section}{1}{\z@}{1.0ex plus
1ex minus .2ex}{.2ex plus .2ex}{\large\bf}}
\def\subsection{\@startsection
   {subsection}{2}{\z@}{.2ex plus 1ex} {.2ex plus .2ex}{\subsize\bf}}
\makeatother

\begin{document}

\date{}

\title{\huge \bf {Sparse Random Approximation and \\ Lossy Compression}}

\author{M. Andrecut 
 \thanks{Manuscript submitted March 21, 2011. Institute for Space Imaging Science, University of Calgary, 
 2500 University Drive NW, Calgary, Alberta, T2N 1N4, Canada, Email: mandrecu@ucalgary.ca.
 }
}

\maketitle
\thispagestyle{empty}


{\hspace{1pc} {\it{\small Abstract}}{\bf{\small--- We discuss a method for sparse signal approximation, 
which is based on the correlation of the target signal with a pseudo-random signal,
and uses a modification of the greedy matching pursuit algorithm.
We show that this approach provides an efficient encoding-decoding
method, which can be used also for lossy compression and encryption
purposes.

\em Keywords: sparse approximation; lossy compression; matching pursuit.}}
 }


\section{Introduction}

Recently there has been an increased interest in alternatives to traditional
signal approximation and compression techniques. Several new methods for
approximating signals in dictionaries of waveforms (Fourier, Gabor,
wavelets, cosine packets etc.) have been proposed \cite{key-1}-\cite{key-5}.
Such a dictionary is a collection of waveforms represented by discrete
time signals, also called atoms. Using this approach, a discrete-time
signal of length $N$ is decomposed in a sparse linear combination
of dictionary atoms with corresponding coefficients. Dictionaries
can be complete, if they contain $N$ atoms, and respectively overcomplete,
if they contain more than $N$ atoms. Most of the dictionaries are
obtained by merging complete dictionaries consisting of different 
types of waveforms. For typical applications the size of these dictionaries
is quite big (tens-of-thousands of atoms), and raises high computational
difficulties and memory requirements. The decomposition of a signal
using overcomplete dictionaries is nonunique, since some elements
in the dictionary have representations in terms of other elements.
Therefore, the sparse approximation problem in an overcomplete dictionary
is to find the minimal representation of a signal, in terms of dictionary
atoms. Finding the sparsest approximation of a signal from an arbitrary
dictionary is an NP-hard problem. Despite of this, several sub-optimal
methods have been recently developed, such that a wide range of applications
(coding, source separation, denoising etc.) have benefited from the
progress made in this area.  

Inspired by the sparse coding paradigm, here we discuss a method for
signal approximation, which is based on the correlation of the target
signal with a pseudo-random signal, and uses a modification of the
greedy matching pursuit algorithm. We show that this approach provides
an efficient encoding-decoding method, with good computational speed and low memory
requirements, which also can be used for lossy compression and encryption
purposes. 


\section{Sparse approximation problem}

The classical signal approximation problem seeks to represent an arbitrary
signal by the best approximation using a restricted class of signals.
The general formulation is as follows \cite{key-6}. Consider $\mathcal{H}$
is an $N-$dimensional Hilbert space (or more generally a Banach space),
with the norm defined in terms of the inner product as:
\begin{equation}
\left\Vert x\right\Vert _{2}=\left\langle x,x\right\rangle ^{1/2}=\left[\sum_{n=0}^{N-1}x_{n}^{2}\right]^{1/2},\;\forall x\in\mathcal{H},
\end{equation}
and an $M$-dimensional subspace $U\subset\mathcal{H}$, $M\leq N$,
spanned by the orthonormal basis: 
\begin{equation}
\{u^{(0)},u^{(1)},...,u^{(M-1)}\}\in U:
\end{equation}
\begin{equation}
\left\langle u^{(i)},u^{(j)}\right\rangle =\delta(i,j)=\left\{ \begin{array}{ccc}
1 & if & i=j\\
0 & if & i\neq j\end{array}\right.,\end{equation}
\begin{equation}
\forall y\in U,\;\exists c_{m},\; m=0,...,M-1,\; y=\sum_{m=0}^{M-1}c_{m}u^{(m)}.
\end{equation}
Given a vector $x\in\mathcal{H}$, the problem is to find the vector
$y\in U$ such that:
\begin{equation}
y=\arg\min_{y\in U}\left\Vert x-y\right\Vert _{2}.
\end{equation}
According to the best approximation theorem, this problem has a unique
solution: 
\begin{equation}
y=\sum_{m=0}^{M-1}c_{m}u^{(m)},
\end{equation}
where
\begin{equation}
c_{m}=\left\langle x,u^{(m)}\right\rangle ,\; m=0,1,...,M-1,
\end{equation}
are the Fourier coefficients. 

The sparse approximation problem is different than the classical one,
since we do not seek a representation in an orthonormal basis, but
on a dictionary $\Phi\in\mathcal{H}$, such that $span(\Phi)=\mathcal{H}$
\cite{key-1}-\cite{key-5}. In general, we consider a countable and
overcomplete dictionary of functions:
\begin{equation}
\Phi=\{\varphi^{(m)}|m=0,1,...,M-1;M\geq N\}
\end{equation}
 in $\mathcal{H}$, which are normalized $(\left\Vert \varphi^{(m)}\right\Vert _{2}=1)$,
but not orthogonal, and possibly redundant. Given $x\in\mathcal{H}$,
the sparse approximation problem consists in finding the sparse coefficients
vector $c\in\mathbb{R}^{M}$, such that:
\begin{equation}
c=\arg\min_{c\in\mathbb{R}^{M}}\left\Vert c\right\Vert _{0},\quad\left\Vert x-\sum_{m=0}^{M-1}c_{m}\varphi^{(m)}\right\Vert _{2}\leq\varepsilon,
\end{equation}
where $\varepsilon\geq0$ is a small positive constant, and:
\begin{equation}
\left\Vert c\right\Vert _{0}=\sum_{m=0}^{M-1}\left[1-\delta(c_{m},0)\right],
\end{equation}
is the $\ell_{0}$ norm, measuring the number of nonzero coefficients.
Obviously, when using an overcomplete dictionary we have more vectors,
and thus a better probability of finding a small number of vectors
that approximate well the given vector. However, since the dictionary
may contain linearly dependent vectors, such an expansion is no longer
unique. Also, in a lossy compression framework the goal is to use
as few vectors as possible, in order to obtain a good approximation.
Unfortunately, this is an NP-hard combinatorial optimization problem,
since in order to find the optimal expansion it is necessary to try
all possible combinations:
\begin{equation}
\left(\begin{array}{c}
M\\
K\end{array}\right)=\frac{M!}{K!(M-K)!},
\end{equation}
searching for the smallest collection of $K$ non-zero terms which
best approximates the signal. 

Several methods have been developed to solve the sparse approximation
problem. The standard approach is based on the convexification of
the objective function, obtained by replacing the $\ell_{0}$ norm
with the $\ell_{1}$ norm \cite{key-1}-\cite{key-4}:
\begin{equation}
\left\Vert c\right\Vert _{1}=\sum_{m=0}^{M-1}\left|c_{m}\right|.
\end{equation}
The resulting optimization problem:
\begin{equation}
c=\arg\min_{c\in\mathbb{R}^{M}}\left\Vert c\right\Vert _{1},\quad\left\Vert x-\sum_{m=0}^{M-1}c_{m}\varphi^{(m)}\right\Vert _{2}\leq\varepsilon,
\end{equation}
is known as Basis Pursuit (BP), and it can be solved using linear
programming techniques whose computational complexities are polynomial
\cite{key-1}. However, in most real applications the BP approach
requires the solution of a very large convex, non-quadratic optimization
problem, and therefore suffers from high computational complexity. 
Another approach is based on greedy algorithms, which are suboptimal
and require far less computations. Our goal is not only to obtain
a good sparse expansion, but also to provide a fast computational
method, therefore here we focus our attention on the greedy Matching
Pursuit (MP) algorithm \cite{key-5}, which is the fastest known algorithm
for the sparse approximation problem. Also, since we are interested
in developing a compression scheme, where only maximum $K\leq N$
out of $M$ dictionary elements can be used in the expansion, we reformulate
the problem as following:
\begin{equation}
c=\arg\min_{c\in\mathbb{R}^{M}}\left\Vert x-\sum_{m=0}^{M-1}c_{m}\varphi^{(m)}\right\Vert _{2},\quad\left\Vert c\right\Vert _{0}\leq K.
\end{equation}


\section{Random dictionary}

Previous studies have shown that the sparsity of the approximation
depends on using an appropriate dictionary for the given class of
signals \cite{key-1}-\cite{key-5}. For example, multiscale decompositions
of natural images into discrete cosine or wavelet bases are quasi-sparse.
Such decompositions have a few significant coefficients, which concentrate
most of the energy and information. This energy compaction property
is then exploited in compression and denoising applications, where
the weak coefficients are usually discarded. Thus, a method to construct
overcomplete dictionaries consists by concatenating orthonormal bases
like: Fourier, Gabor functions, wavelets, cosine packets etc. These
dictionaries can be improved by employing learning methods \cite{key-7},
which adapt an initial dictionary to a set of training samples. In
this case the goal is to optimize a dictionary, such that a given
class of signals, has a sparse approximation. 

The sparse decomposition
abilities of such dictionaries are characterized by the restricted
isometry property (RIP) of the $N\times M$ matrix $\hat{\Phi}$,
with the columns given by the dictionary atoms $\varphi^{(m)}$.
The $K$-restricted isometry constant $\delta_{K}$ of  $\hat{\Phi}$ is the smallest
quantity such that for every $K$-sparse vector $c\in\mathbb{R}^{M}$
we have \cite{key-2}-\cite{key-4}:
\begin{equation}
(1-\delta_{K})\left\Vert c\right\Vert _{2}^{2}\leq\left\Vert \hat{\Phi}c\right\Vert _{2}^{2}\leq(1+\delta_{K})\left\Vert c\right\Vert _{2}^{2}.
\end{equation}

This means that every set of less than $K$ columns are approximately
orthogonal. Smaller $\delta_{K}$ means better orthogonality, and
therefore a better discrimination capability of atoms. Recently it
has been shown that random matrices satisfy the RIP with high probability
\cite{key-2}-\cite{key-4}. Therefore, some good examples of overcomplete dictionaries
include: 

- matrices of independent and identical distributed (i.i.d) Gaussian samples from $N(0,1)$; 

- matrices with Bernoulli entries, where $\varphi_{nm}=\pm1$ with equal probability $p=1/2$;

- matrices with randomly sampled Fourier elements etc.

Inspired by the above results obtained for random matrices, here we
propose a simplified approach. Instead of using a large random dictionary,
we simply use a random vector of length $N+M$:
\begin{equation}
f=[f_{0},f_{1},...,f_{M-1+N}]^{T}\in\mathbb{R}^{N+M},
\end{equation}
generated by a reproducible random process, a pseudo-random number
generator for example. An atom of this dictionary will simply
be a normalized ``window'' vector:
\begin{equation}
\varphi^{(m)}=\frac{[f_{m},f_{m+1},...,f_{m+N-1}]^{T}}{\sqrt{\sum_{n=0}^{N-1}f_{m+n}^{2}}}\in\mathbb{R}^{N}.
\end{equation}
Obviously, there are $M$ such window vectors in any realization of
the pseudo-random process, and they are uncorrelated since $f_{m}$
are i.i.d random variables (depending on the quality of the random
number generator used). Such a vector can be easily generated by encoder
and decoder using only a given seed value. This way we avoid the high
memory and computational requirements, while still obtaining a good
sparse decomposition, which we will show that it can be also efficiently
used for lossy compression. Without loosing generalization, in order
to obtain an easy normalization we consider that each $f_{m}$
is a Bernoulli random variable. Thus, the normalization is easily
achieved by simply dividing the vector $f$ with $\sqrt{N}$. 
We should mention that the mean $\left\langle x\right\rangle $ of the signal 
can be captured correctly if we assume that the components of first
atom of the dictionary are all set to $1$: 
\begin{equation}
\varphi^{(0)}=\frac{1}{\sqrt{N}}[1,1,...,1]^{T}.
\end{equation}

This dictionary choice also provides a simple encryption scheme, since
for different seeds one obtains different dictionaries. Therefore, if the
seed is user defined then the obtained expansion will also be encrypted,
since the same secret seed is needed for decoding.

\section{Matching pursuit}

Matching Pursuit is a well known greedy algorithm widely used in approximation
theory and statistics \cite{key-5}. One of its main features is that
it can be applied to arbitrary dictionaries. Starting from an initial
approximation $c=0$ and residual $r=x$, the algorithm uses an iterative
greedy strategy to pick the dictionary atoms that are the most
strongly correlated with the residual. Then, successively their contribution
is subtracted from the residual, which this way can be made arbitrarily
small. Using the simplified dictionary $f$, the pseudo-code of the
MP algorithm takes the form listed in Algorithm 1.

--------------------------------------------------------------------------

Algorithm 1. Matching Pursuit (MP)

--------------------------------------------------------------------------

$K$; // number of atoms in the approximation 

$c\leftarrow0$; // coefficients of selected atoms 

$p\leftarrow0$; // positions of selected atoms

$r\leftarrow x$; // initial residual

for($k=0,1,...,K-1$)\{

\quad{}$s_{max}\leftarrow0$;

\quad{}for($m=0,1,...,M-1$)\{

\quad{}\quad{}$s\leftarrow\left\langle r,\varphi^{(m)}\right\rangle $;

\quad{}\quad{}if($\left|s\right|>\left|s_{max}\right|$)\{

\quad{}\quad{}\quad{}$s_{max}\leftarrow s$;

\quad{}\quad{}\quad{}$i\leftarrow m$;\}\}

\quad{}$p_{k}\leftarrow i$;

\quad{}$c_{k}\leftarrow s_{max}$;

\quad{}$r\leftarrow r-c_{k}\varphi^{(p_{k})}$;\}

return $p$, $c$;

--------------------------------------------------------------------------

Thus, at each iteration step $k=0,1,...,K-1$ the algorithm selects
the index $p_{k}$ of the atom $\varphi^{(p_{k})}$, which has
the highest correlation with the current residual, and updates the
estimate of the corresponding coefficient $c_{k}$, and the residual
$r$. After $K$ selection steps the algorithm returns the positions
$p$ of the selected atoms in the dictionary and their corresponding
coefficients $c$ in the expansion. A shortcoming of the MP algorithm
is that although the asymptotic convergence is guaranteed and it can
be easily proved, the resulting approximation after any finite number
of steps $K\leq N$ will in general be suboptimal. Thus, one cannot
expect an exact reconstruction of the target signal after decoding. 

One can see that the decoding step requires both the positions $p$
and the coefficients $c$ of the selected atoms in order to compute the 
$K$-term approximation:
\begin{equation}
y\leftarrow\sum_{k=0}^{K-1}c_{k}\varphi^{(p_{k})}\simeq x.
\end{equation}
This is inconvenient from the point of view of compression. Assuming
for example that the elements of the input vector $x\in\mathbb{R}^{N}$
are floating point numbers represented on $Q$ bits, we need $NQ$
bits to store the whole vector. Thus, in order to compress $x$ we
need to reduce this number. For each position we need $Q$ bits, and
for each coefficient we also need $Q$ bits, therefore the output
of the MP algorithm requires $2KQ$ bits. Thus, in order to achieve
compression we must have $K<N/2$. This condition can be relaxed by
imposing that both $p_{k}$ and $c_{k}$ are stored together as a
single number represented on $Q$ bits. Of course, one may think that
in this case the precision of $c_{k}$ will be affected, and the approximation
will deteriorate significantly. However, due to the random characteristic
of the dictionary we may expect that this may actually work, and at
each step the algorithm will pick the atom with the best pair $(p_{k},c_{k})$ which
can be ``accommodated'' on a number $h_{k}$ represented on $Q$
bits. Thus, instead of using $2Q$ bits to store a pair $(p_{k},c_{k})$,
we may actually use only $Q$ bits to store their equivalent $h_{k}$.
The question is how to do this efficiently? 

We observe that we need to find the atom characterized by a pair
$(p_{k},c_{k})$, where there is a trade-off between the necessary
precision of $c_{k}$ and the length of $p_{k}$, such that they can
be represented together on $Q$ bits, and their inclusion in the approximation
expansion decreases the residual $r$. Ideally, we would like to allocate
$Q/2$ bits for the position $p_{k}$, and $Q/2$ bits for the corresponding
coefficient $c_{k}$. That means to restrict the length of the dictionary
to $M=2^{Q/2}$. Thus, for a typical integer representation on $Q=32$
bits, the dictionary will contain $M=2^{16}$ elements. 

Now, let us assume the signal is normalized $x\leftarrow x/\left\Vert x\right\Vert $
before the compression. Since the MP algorithm will always produce
a residual $r$ with $\left\Vert r\right\Vert <\left\Vert x\right\Vert =1$,
the result of the correlation term $s\leftarrow\left\langle r,\varphi^{(m)}\right\rangle $,
will always be bounded: $s\in(-1,1)$. Thus, the value of a resulted
coefficient will also be bounded by the same interval: $c_{k}\in(-1,1)$.
Finally, the idea is to make the position $p_{k}$ equal with the
integer part of $h_{k}$, and the coefficient $c_{k}$ equal with the
fractional part of $h_{k}$:
\begin{equation}
int(h_{k})\leftarrow p_{k},\quad frac(h_{k})\leftarrow c_{k}.
\end{equation}
Obviously, by doing this some of the precision in the representation
of $c_{k}$ will be lost. The compressive MP
(CMP) algorithm which takes into account these modifications is listed in Algorithm 2.

One can see that the computation of the cross-correlation term $s$
requires two extra steps. In the first step, the index $m$ of the
currently tested atom $\varphi^{(m)}$, and its correlation $s\in(-1,1)$
with the residual, are packet together in $s$, using: 
\begin{equation}
s\leftarrow sign(s)m+s.
\end{equation}
This will result in a loss of the precision of the fractional part
of $s$, since it needs to accommodate also the integer part $m$,
on the same number $Q$ of bits. In the second step we extract the
resulted fractional part using: 
\begin{equation}
s\leftarrow sign(s)(\left|s\right|-\left|int(s)\right|),
\end{equation}
in order to perform the comparison with the current maximum value
$\left|s_{max}\right|$. If the test is true, then we store the obtained
values in $s_{max}\leftarrow s$, and respectively $i\leftarrow m$.
The values $s_{max}$ and $i$, corresponding to the the best atom 
are then used to update the residual, and they are packed into $h_{k}$,
this time without information loss. Thus, after the first packing
step, when some precision is lost, the future packing-unpacking steps
become reversible, and the precision is conserved. 

--------------------------------------------------------------------------

Algorithm 2. Compressive Matching Pursuit (CMP)

--------------------------------------------------------------------------

$K$; // number of atoms in the approximation 

$h\leftarrow0$; // positions and coefficients of selected atoms

$h_{K}\leftarrow\left\Vert x\right\Vert _{2}$; // signal normalization

$r\leftarrow x/h_{K}$; // initial residual

for($k=0,1,...,K-1$)\{

\quad{}$s_{max}\leftarrow0$;

\quad{}for($m=0,1,...,M-1$)\{

\quad{}\quad{}$s\leftarrow\left\langle r,\varphi^{(m)}\right\rangle $;

\quad{}\quad{}$s\leftarrow sign(s)m+s$;

\quad{}\quad{}$s\leftarrow sign(s)(\left|s\right|-\left|int(s)\right|)$;

\quad{}\quad{}if($\left|s\right|>\left|s_{max}\right|$)\{

\quad{}\quad{}\quad{}$s_{max}\leftarrow s$;

\quad{}\quad{}\quad{}$i\leftarrow m$;\}\}

\quad{}$r\leftarrow r-s_{max}\varphi^{(i)}$;

\quad{}$h_{k}\leftarrow sign(s_{max})i+s_{max}$;\}

return $h$, $\varepsilon$;

--------------------------------------------------------------------------

We should also save the norm of $x$, which is required for the decoding step: 
$h_{K}=\left\Vert x\right\Vert _{2}$.
Therefore, the length of the vector $h$ is $K+1$, where the first $K$
values correspond to the positions (the positive integer part) and
coefficients (the fractional part) of the atoms selected in the
approximation expansion, while the last value contains the norm
of the input signal. The decoding procedure is very fast and it is done as following:
\begin{equation}
y\leftarrow h_{K}\sum_{k=0}^{K-1}sign(h_{k})(\left|h_{k}\right|-\left|int(h_{k})\right|)\varphi^{(\left|int(h_{k})\right|)} \simeq x.
\end{equation}

\begin{figure}
\centering
\includegraphics[scale=1]{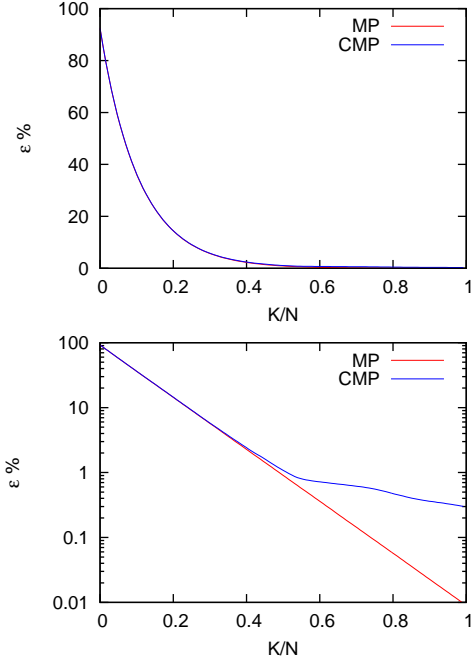}
\caption{\label{Fig1} Relative approximation error $\epsilon$ of the MP and CMP algorithms.}
\end{figure}

\section{Numerical results}

We have implemented the MP and CMP algorithms in parallel using C,
OpenMP and GCC on a Linux platform. The parallel OpenMP version is almost $n$ times 
faster than the serial version, where $n$ is the number of available CPU cores. 

In the following experiments we consider a random
Bernoulli dictionary with $M=2^{16}$ elements. Also, in order to
speed up computation we set the length $N$ of the input signal to
$N=128$. A longer input signal can be easily divided in chunks of
length $N\leq 128$, which can be then independently processed. Ideally,
we would like to approximate and compress any kind of signal, so the
shape of the signal doesn't really matter. Therefore, for testing
we choose random samples drawn from a uniform distribution on the
interval (-2$^{15}$,2$^{15}$), represented as floating point numbers
on $Q=32$ bits. This, means that we practically approximate and compress
noise. This is usually a difficult task, since such signals will have
the widest possible bandwidth, for the considered finite length $N$
of the samples. The quantity of interest is the relative recovery
error, which is defined as following: 
\begin{equation}
\varepsilon=100\frac{\left\Vert x-y\right\Vert _{2}}{\left\Vert x-\left\langle x\right\rangle \right\Vert _{2}},
\end{equation}
where $\left\langle x\right\rangle $ is the signal mean value, and
$y$ is the $K$-term approximation expansion. In Figure 1 we give
the value of $\varepsilon$ as a function of the inverse of compression
ratio $\rho^{-1}=K/N$ (linear scale - top, logarithmic scale - bottom).
The results were averaged over 1000 samples for $K=1,...,N$. One
can see that the effect of the CMP packing procedure 
$(p,c)\rightarrow h$ manifests only for $K/N>0.5$, which
is obviously not too bad for lossy compression purposes. Also, we
have $\varepsilon(0.5)\simeq1\%$, which means that a compression
ratio of $\rho_{CMP}=2:1$ produces a distortion of the data of only
$1\%$. 

\begin{figure}
\centering
\includegraphics[scale=0.7]{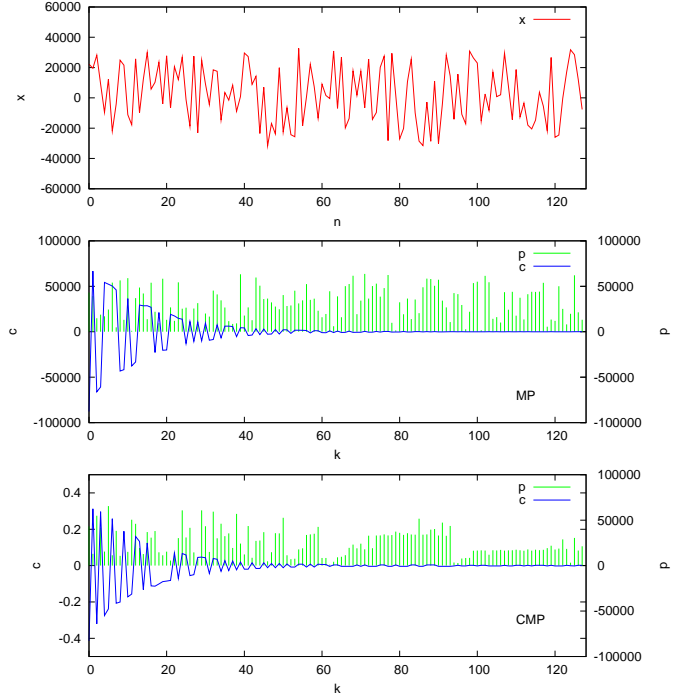}
\caption{\label{Fig2} Decomposition of a random signal using MP and CMP algorithms.}
\end{figure}

\begin{figure}
\centering
\includegraphics[scale=0.7]{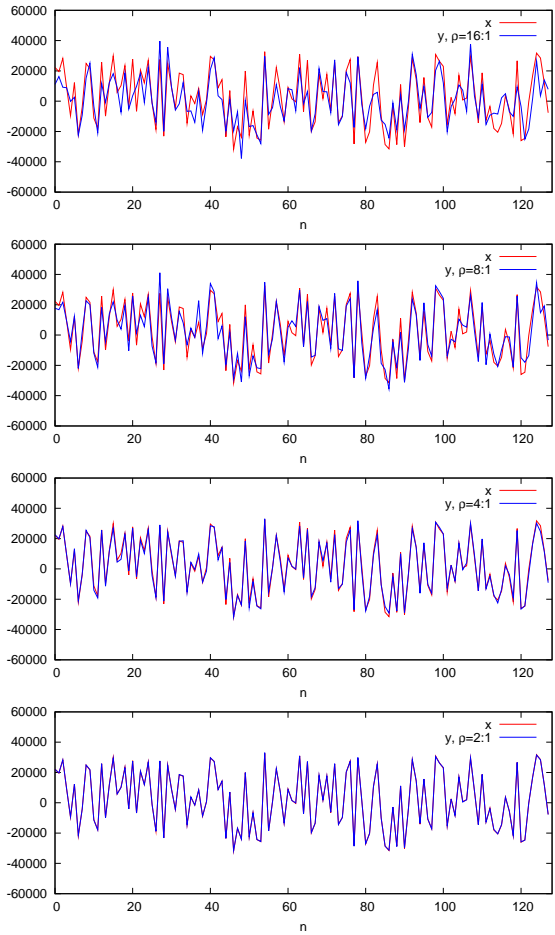}
\caption{\label{Fig3} Lossy compression of a random signal.}
\end{figure}

\begin{figure}
\centering
\includegraphics[scale=0.7]{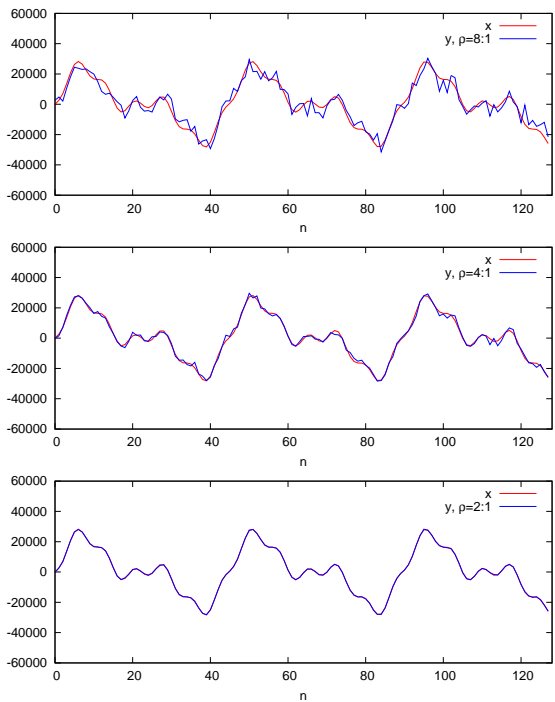}
\caption{\label{Fig4} Lossy compression of a smooth signal.}
\end{figure}

In Figure 2 we give a typical sample, and its decomposition
by the MP and CMP algorithms. Here we have represented the values
of the coefficients $c_{k}$ and the positions $p_{k}$ of the corresponding
atoms in the dictionary. In the case of CMP, the positions and
the coefficients are extracted from $h_{k}$, using: $p_{k}=int(h_{k})$,
and $c_{k}=sign(h_{k})(\left|h_{k}\right|-\left|p_{k}\right|)$.
The positions and the coefficients for MP and CMP are different, due
to the extra packing constraint in the CMP algorithm. Also, we should
notice the exponential decrease of the magnitude of the coefficients
$c_{k}$. Thus, a lossy compression of the signal can be achieved
retaining only the first coefficients with large values. 

In Figure 3 we have the same signal from Figure 2, and its recovery after lossy compression, 
for several different compression ratios: $\rho_{CMP}=16:1$, $\varepsilon=51.30\%$;
$\rho_{CMP}=8:1$, $\varepsilon=28.25\%$; $\rho_{CMP}=4:1$, $\varepsilon=9.14\%$;
and respectively $\rho_{CMP}=2:1$, $\varepsilon=1.01\%$. Similar results
have been obtained for different types of signals. For example, in
Figure 4 we consider a smooth signal (a superposition of sinusoids) for several compression ratios:
$\rho_{CMP}=8:1$, $\varepsilon=27.83\%$; $\rho_{CMP}=4:1$, $\varepsilon=8.92\%$;
and respectively $\rho_{CMP}=2:1$, $\varepsilon=0.98\%$.

\section*{Conclusion}

We have discussed a sparse random approximation method, based on the
correlation of the target signal with a pseudo-random signal, and
a modification of the greedy matching pursuit algorithm. We have
shown that this approach provides an efficient encoding-decoding method.
Also, the presented method has the advantage of an easy implementation,
with high computational speed and low memory requirements, which also
can be used for lossy compression and encryption purposes.


\end{document}